\documentclass{aa}  
\usepackage{color}
\usepackage{comment}
\usepackage{natbib}
\usepackage{url}
\usepackage{subfigure}
\usepackage{xspace}
\usepackage[percent]{overpic}
\usepackage{subfloat}

\usepackage{amsmath,amstext}
\usepackage{balance}
\usepackage{booktabs}
\usepackage{multirow}
\usepackage[colorlinks=true, linkcolor=blue, citecolor=blue, urlcolor=cyan]{hyperref}
\usepackage{orcidlink}

\newcommand{\nustar}{{\it NuSTAR}\xspace}

\newcommand{\swift}{{\it Swift}\xspace}
\newcommand{\xmm}{{\it XMM-Newton}\xspace}

\newcommand{\src}{MAXI~J0655}

\graphicspath{{./}{figures/}}

\begin{document} 
   \title{Double-hump spectrum, pulse profile dip, and pulsed fraction spectra from the low-accretion regime in the X-ray pulsar \src-013}
    \titlerunning{\src-013 observed at low luminosity}
\authorrunning{Malacaria et al.}

\author{
C.~Malacaria\orcidlink{0000-0002-0380-0041}\inst{\ref{in:INAF-OAR}}
\and 
S.~N.~Pike \inst{\ref{in:UniCal}}
\and
A.~D'Aì\orcidlink{0000-0002-5042-1036}\inst{\ref{in:iasfpa}}
\and G.~L.~Israel\orcidlink{0000-0001-5480-6438}\inst{\ref{in:INAF-OAR}}
\and L.~Ducci\orcidlink{0000-0002-9989-538X} \inst{\ref{in:UniTue}}
\and
R.~E.~Rothschild\inst{\ref{in:ucsd}}
\and L.~Stella\inst{\ref{in:INAF-OAR}}
\and R.~Amato\inst{\ref{in:INAF-OAR}}
\and E.~Ambrosi \inst{\ref{in:iasfpa}} 
\and J.~B.~Coley
\orcidlink{0000-0001-7532-8359} \inst{\ref{in:Howard},\ref{in:CRESST}}
\and F.~F\"urst 
\orcidlink{0000-0003-0388-0560}
\inst{\ref{in:esa}}
\and M.~Imbrogno\orcidlink{0000-0001-8688-9784}\inst{\ref{in:ICE},\ref{in:IEEC},\ref{in:INAF-OAR}}
\and P.~Kretschmar\orcidlink{0000-0001-9840-2048}\inst{\ref{in:esa}}
\and D.~K.~Maniadakis
\orcidlink{0009-0008-1148-2320} 
\inst{\ref{in:iasfpa},\ref{in:unipa}}
\and
A. Papitto\inst{\ref{in:INAF-OAR}}
\and P.~Pradhan\orcidlink{0000-0002-1131-3059}\inst{\ref{in:erau}}
\and A.~Rouco~Escorial\orcidlink{0000-0003-3937-0618}\inst{\ref{in:esa}}
\and A.~Simongini 
\orcidlink{0009-0000-3416-9865} \inst{\ref{in:INAF-OAR},\ref{in:tov}}
\and J.~Stierhof\orcidlink{0000-0001-9537-7887}\inst{\ref{in:bamberg}}
\and B.~F.~West\orcidlink{0009-0007-1219-7508}\inst{\ref{in:USNA}}
\and N.~Zalot\inst{\ref{in:bamberg}}
}

\institute{
INAF-Osservatorio Astronomico di Roma, Via Frascati 33, I-00078, Monte Porzio Catone (RM), Italy \label{in:INAF-OAR} \\ \email{christian.malacaria@inaf.it}
\and Department of Astronomy and Astrophysics, University of California, San Diego, CA 92093, USA \label{in:UniCal}
\and INAF - IASF Palermo, via Ugo La Malfa 153, 90146 Palermo, Italy \label{in:iasfpa}
\and
Department of Astronomy and Astrophysics, University of California San Diego, La Jolla, CA 92075, USA \label{in:ucsd}
\and Institut f\"ur Astronomie und Astrophysik, Universit\"at T\"ubingen, Sand 1, D-72076 T\"ubingen, Germany\label{in:UniTue}
\and Institute of Space Sciences (ICE, CSIC), Campus UAB, Carrer de Can Magrans s/n, E-08193, Barcelona, Spain\label{in:ICE}
\and Institut d'Estudis Espacials de Catalunya (IEEC), 08860 Castelldefels (Barcelona), Spain\label{in:IEEC}
\and European Space Agency (ESA), European Space Astronomy Centre (ESAC), Camino Bajo del Castillo s/n, 28692 Villanueva de la Cañada, Madrid, Spain \label{in:esa}
\and Università Tor Vergata, Dipartimento di Fisica, Via della Ricerca Scientifica 1, I-00133 Rome, Italy \label{in:tov}
\and Department of Physics and Astronomy, Howard University, Washington, DC 20059, USA\label{in:Howard}
\and CRESST and NASA Goddard Space Flight Center, Astrophysics Science Division, 8800 Greenbelt Road, Greenbelt, MD, USA\label{in:CRESST}
\and Dipartimento di Fisica e Chimica E. Segrè, Università degli Studi di Palermo, Via Archirafi 36, 90123 Palermo, Italy\label{in:unipa}
\and Physics Department, United States Naval Academy, Annapolis, MD 21402, USA \label{in:USNA}
\and Department of Physics and Astronomy, Embry-Riddle Aeronautical University, 3700 Willow Creek Road, Prescott, AZ, USA, 86301 \label{in:erau}
\and Dr.\ Karl-Remeis-Observatory and Erlangen Centre for Astroparticle Physics, Friedrich-Alexander-Universit\"at Erlangen-N\"urnberg, Sternwartstr.~7, 96049 Bamberg, Germany \label{in:bamberg}}

\date{\today}
% \abstract{}{}{}{}{} 
% 5 {} token are mandatory
  \abstract
  % context heading (optional)
  % {} leave it empty if necessary  
   {Accreting X-ray pulsars (XRPs) undergo different physical regimes depending on different mass accretion rates, with related changes in the emitted X-ray spectral properties. 
   Recent observations have shown a dramatic change in the emission properties of this class of sources observed at low luminosity.}
  % aims heading (mandatory)
   {We explore the timing and spectral properties of the XRP \src-013 observed in the low-luminosity regime (about $5\times10^{33}\,$erg\,s$^{-1}$) to witness the corresponding spectral shape and pulse profiles.
   }
   {We employ recent \xmm\ and \nustar\ pointed observations of the \src-013 X-ray activity    during the low-luminosity stage.
   We explore several spectral models to fit the data and test theoretical expectations of the dramatic transition of the spectral shape    compared to the higher luminosity regime.
   We study the pulsating nature of the source and find a precise timing solution. We explore the energy-resolved pulse profiles and the derived energy-dependence of different pulsed fraction estimators ($\mathrm{PF_\mathrm{minmax}}$ and $\mathrm{PF_\mathrm{rms}}$).
   We also obtain \nustar\ pulsed fraction spectra (PFS) at    different luminosity regimes.}
  % results heading (mandatory)
   {\src-013 spectrum is well fitted by a double Comptonization model, in agreement with recent observational results and theoretical expectations that explain the observed spectrum as being composed of two distinct bumps, each dominated by different polarization modes. 
   We measure a spin period of $1081.86\pm0.02\,$s, consistent with the source spinning-up compared to previous observations, yielding an upper limit for the magnetic field strength of $B\lesssim9\times10^{13}\,$G.
   The pulse profiles %at all investigated energies 
   show a single broad peak interrupted by a sharp dip that coincides with an increase in the hardness ratio, thus likely due to absorption. For the low-luminosity observation, the $\mathrm{PF_\mathrm{minmax}}$ increases with energy up to $\sim100\%$ in the 10-30 keV band, while the $\mathrm{PF_\mathrm{rms}}$ remains steady at $\sim60\%$. The PFS obtained at high luminosity shows evidence of an iron $K\alpha$ emission line but no indications of a cyclotron line.
   }
  {}
  % conclusions heading (optional), leave it empty if necessary 
  
   \keywords{X-ray binary stars -- stars: neutron -- pulsars: individual: \src-013 -- accretion, accretion disks, polarization -- magnetic fields
               }
   \maketitle
%
%-------------------------------------------------------------------
%\vspace{-10.}
\section{Introduction}

MAXI~J0655-013 (J0655 hereafter) is an accreting X-ray pulsar (XRP) recently discovered \citep{Serino22, Kennea22, Nakajima22} during an outburst episode that allowed its spectral and timing broadband study with \nustar\ (\citealt{Pike2023}, and references therein). Besides that one outburst episode, the source has not shown any evident X-ray activity. The optical companion is a B1-3e III-V star, implying that the system is a Be/X-ray Binary \citep[BeXRB,][]{Reig22_atel}. The pulsar is a slowly rotating neutron star (NS) with a spin period of about 1100 s \citep{Shidatsu22, Pike2023}. 
The broadband \nustar\ continuum spectrum of the source was fitted with the typical model employed for this class of sources, i.e., a cutoff power-law model \citep{Pike2023}.
The magnetic field strength for this source has been estimated to be in the range $0.5 - 50\times10^{13}$ G \citep{Pike2023}.
The wide range of magnetic field values derives from the different methods employed to infer it, including a tentative cyclotron line at 44 keV. The line was, however, interpreted with caution by the authors for several reasons, e.g., the width of the line is too narrow to be consistent with other similar sources and accretion conditions.
A distance of $3.6^{+0.3}_{-0.2}\,$kpc was measured by Gaia DR3 \citep{Gaia2023}. Given this distance value, \citet{Pike2023} measured the source X-ray luminosity at the early and late stages of the 2022 outburst as $5.6\times10^{36}\,$erg\,s$^{-1}$ and $4\times10^{34}\,$erg\,s$^{-1}$, respectively.

The spectral energy distribution of accreting XRPs observed during bright ($\sim10^{36}\,$erg\,s$^{-1}$) outburst episodes is typically fitted using a phenomenological cutoff power-law model \citep{Nagase89,Caballero+Wilms12, Mushtukov22}. While this approach is a useful first approximation, it does not provide insights into the ongoing accretion physics. In fact, the computation of physical parameters needs to account for magnetic, thermal, and bulk Compton scattering, cyclotron emission and absorption, blackbody radiation from multiple regions, and magnetic bremsstrahlung \citep[and references therein]{Becker2022}.  
However, recent studies of XRPs observed at low-mass accretion rates (thus low luminosity values, i.e., $\sim10^{33}-10^{34}\,$erg\,s$^{-1}$) have highlighted that the spectrum observed in this regime shows a substantially different shape.
In fact, the spectral distribution observed at low luminosity \textbf{often} shows two main distinct components, indicated as a double-hump spectrum \citep{Tsygankov2019, Tsygankov2019_a0535, Lutovinov21}. This is interpreted as the result of different mechanisms occurring at spatially separate sites, as follows. The upper layers of the atmosphere are overheated (compared to the lower layers) by residual accretion. At the same time, seed blackbody photons are produced by magnetic free-free emission in the lower part of the atmosphere near the NS surface. While collisions cause the braking of accretion in the atmosphere, resulting in the excitation of electrons to upper Landau levels, the subsequent radiative de-excitation of electrons produces cyclotron photons.
Cyclotron and free-free photons reprocessed by resonant Comptonization in an atmosphere with a strong temperature gradient are responsible for the high-energy bump peaking around 30 keV, while the absorbed part is then released mostly as extra-ordinary polarized photons from the bottom part of the atmosphere, thus forming the low-energy bump peaking around 5 keV \citep{Mushtukov2021, Sokolova-Lapa2021}.

In this work, we take advantage of pointed X-ray observations of the accreting XRP \src\ in order to study its low-luminosity spectral and timing properties. Due to its residual accretion, its vicinity, an estimation of its magnetic field, and a relatively unobscured line of sight, \src\ is an ideal target to unveil the processes responsible for the characteristic spectral shape and the accompanying timing features in the low-luminosity regime. 

%--------------------------------------------------------------------
\section{Observations and data reduction}

\begin{figure*}[!t]
    \centering
    \includegraphics[width=\textwidth]{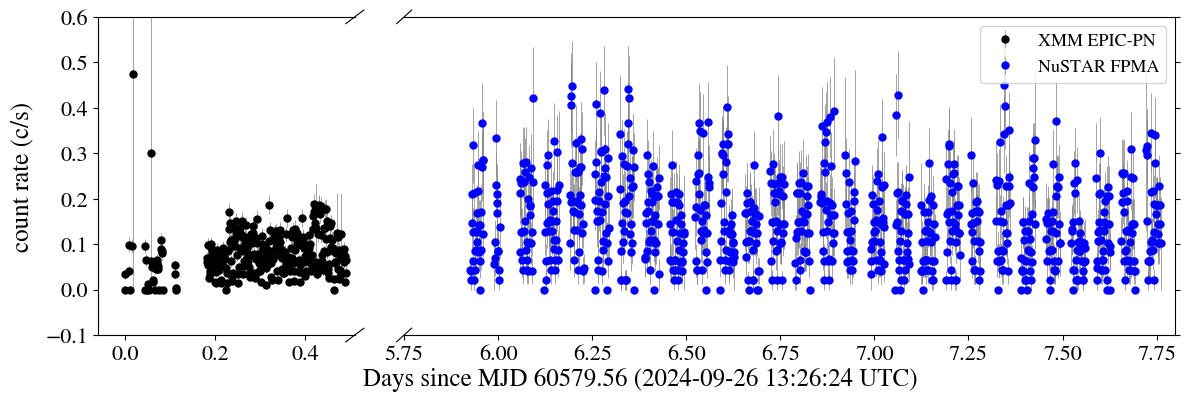}
    \caption{\xmm\ (0.5-10 keV) and \nustar\ (4-79 keV) both at 100\,s binning light curves of \src. Count rates were rescaled according to model predictions and effective area accounting for the observed spectral energy distribution.}
    \label{fig:lightcurve}
\end{figure*}

A light curve of \src\ as observed by our pointed \xmm\ and \nustar\ observations is shown in Fig.~\ref{fig:lightcurve}. Observations are separated by a 1-week gap. In the following, we illustrate the data reduction procedure for each dataset.

\subsection{\xmm\ observations}\label{subsec:xmm-analysis}

\xmm\ observed \src\ on September 26, 2024 (MJD 60579, see Table~\ref{table:log}). ObsID 0935630101 was carried out in Full Frame mode.
\xmm\ data were reduced using the \xmm\ Science Analysis System (SAS) v21.0.0 and the latest available calibration files (XMM-CCF-REL-403). We followed the standard procedure described in SAS Data Analysis Threads for data reduction\footnote{\url{https://www.cosmos.esa.int/web/XMM-Newton/sas-threads.}}. Due to high solar activity, the effective exposure was reduced from the nominal 66 ks to about 40 ks for EPIC-MOS and 25 ks for EPIC-pn.

After standard cleaning of the raw data, events were extracted from a circular region centered on the source with a radius of 24\arcsec, while background events were extracted from a 75\arcsec-radius circular source-free region on the same chip. Spectral response files were generated using the \texttt{rmfgen} and \texttt{arfgen} tasks. 
We also applied a correction to the effective area suited for simultaneous spectral fitting of EPIC and \nustar\ spectra available through the \texttt{applyabsfluxcorr=yes} option.
We verified through the \textit{epatplot} tool that pile-up did not affect our spectra.

Finally, we also extracted EPIC-pn and EPIC-MOS spectra for GTIs where the flaring particle background was high.
We then compared these spectra with those obtained from events outside the flaring activity period.
The spectra obtained from the flaring part of the data were fully consistent with those obtained by filtering out the high-flaring period. Therefore, we merged the resulting spectra and responses using the \texttt{epicspeccombine} tool, increasing the exposures by about 6 ks and 16 ks for EPIC-MOS and EPIC-pn, respectively. 
We also verified that both in the case when individual spectra are used for the \nustar\ joint fit, and in the case when the \xmm\ merged spectra are used, the same cross-calibration constant value is obtained (see Sect. \ref{sec:spectral_analysis}).

\subsection{NuSTAR}

\nustar observed \src\ on October 10, 2024
(MJD 60593, ObsID 31001020002). \nustar data were reduced with \texttt{NUSTARDAS} v2.1.2 using the \texttt{CALDB} 20240729 \citep{Madsen21}. The cleaned events were obtained following the standard \nustar guidelines. Source spectra were extracted through the \texttt{NUPRODUCTS} routine. The source extraction region was a circular region of 50\arcsec\ centered on the source, while background events were extracted from a 90\arcsec\ circular region of a source-free area on the same detector.
Due to increased Solar activity during observation, an additional background filtering based on the South Atlantic Anomaly (SAA) tentacles and space weather conditions was applied (\texttt{saacalc=2 saamode=OPTIMIZED tentacle=yes}). 
However, we verified that such strict filtering does not yield an improvement in the flaring background, at the expense of a considerable part of the exposure, compared to the employment of the \texttt{OPTIMIZED} option for SAA filtering alone. Therefore, we opted for the latter filtering.
The resulting total filtered exposure time was approximately 83 ks.

\subsection{Swift/XRT}

Upon further research of archival data, the source was found to be already known from the Improved and Expanded version of the Neil Gehrels Swift X-ray Telescope Point Source Catalog as 2SXPS J065512.4-012855 \citep{Evans2020}, where it was serendipitously cataloged as an X-ray source. 
\swift/XRT observed this source on May 23, 2018 (MJD 58261, ObsID 03101816001).
The source was observed in photon counting mode for a total of about 1 ks (see Table\,\ref{table:log}).
We extracted XRT data products using the online platform \citep{Evans2009} provided by the UK \swift\ Data Centre\footnote{\url{http://www.swift.ac.uk/user_objects/}}.

\begin{table}[!t]
\caption{Log of \src\ observations used in this work.} \label{table:log}
\centering
\begin{tabular}{lcc}
\hline\hline 
 ObsID & Time & Exposure\\
& &  [ks]  \\
\xmm\\
0935630101 & 2024-09-26 & 66\\
\\
\textit{Nustar}\\
31001020002 & 2024-10-02 & 84\\
\\
\textit{Swift}/XRT\\
03101816001 & 2018-05-23 & 0.95\\
\midrule
\hline 
\end{tabular}
%\tablefoot{}
\end{table} 

\section{Spectral analysis}\label{sec:spectral_analysis}

We analyzed spectral data using \texttt{XSPEC} \texttt{v12.13.1} \citep{Arnaud96} as part of \texttt{HEASOFT v6.33.1}.
\xmm\ spectral data were rebinned using the \texttt{specgroup} routine with the standard value of 25 counts per bin and an oversampling factor of the intrinsic energy resolution of up to 3, while \nustar\ spectra were rebinned with the \texttt{ftgrouppha} \texttt{HEASARC} tool with the \citet{Kaastra2016} optimal binning and the additional requirement of a minimum number of 25 counts per bin. We employ $\chi^2$ as the test statistic, and Cash statistic \citep{Cash79} with Poissonian background, that is, W-stat, as the fit statistic. 
The photoelectric absorption component (or column density N$_{\rm H}$) was set according to \citet[\texttt{tbabs} in XSPEC]{Wilms00}, and we assumed model-relative (\texttt{wilm}) elemental abundances.

\begin{figure}[!t]
    \centering
    \includegraphics[width=\columnwidth]{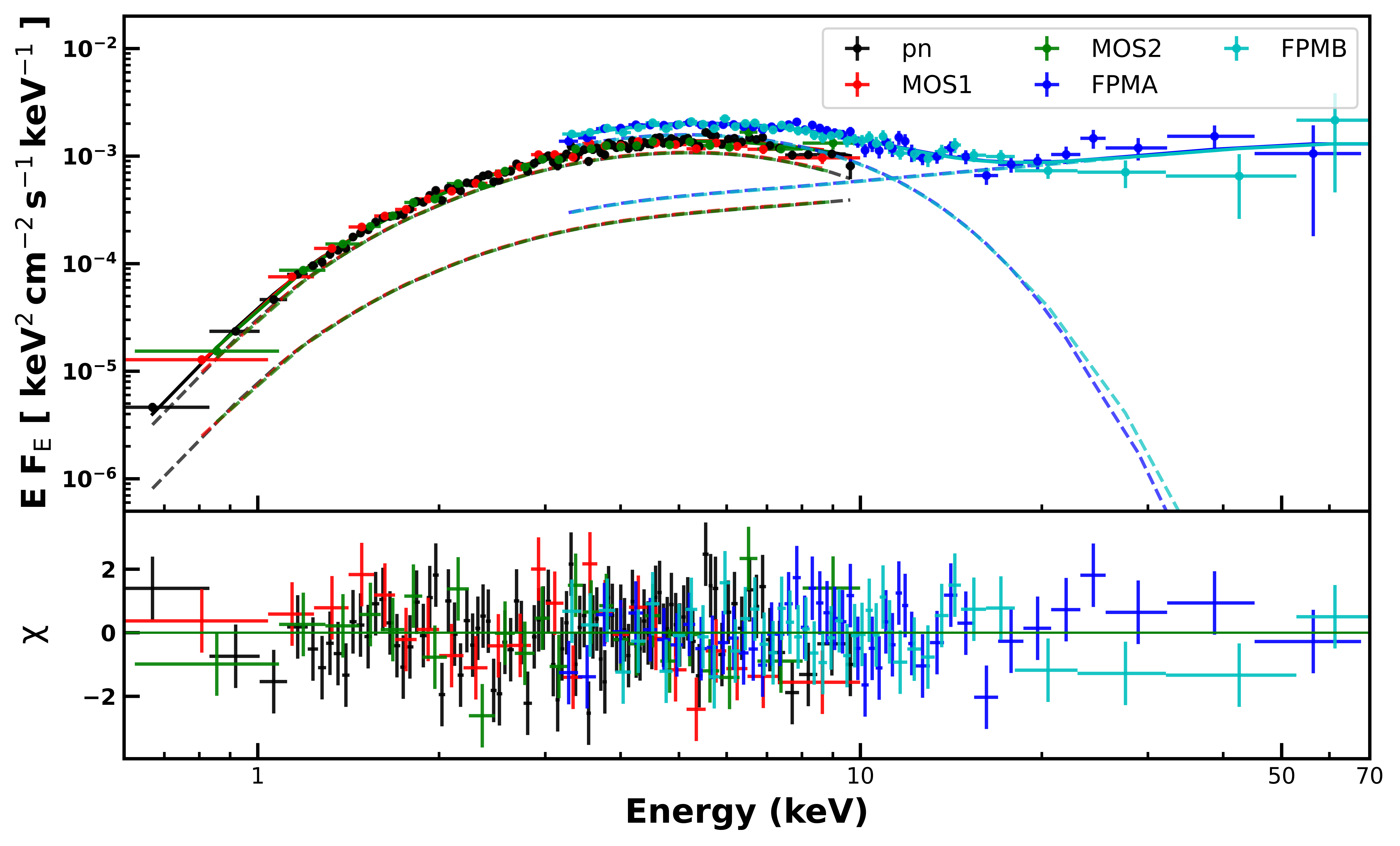}\label{fig:double-hump_spectrum}
    \caption{Unfolded spectra from \xmm\ pn, MOS 1 and 2 (black, red and green, respectively) and \nustar\ FPMA and FPMB (blues and cyan, respectively) from the \src\ observations in the low-luminosity regime analyzed in this work. Dashed lines represent the two Comptonization components for each instrument. Bottom panel: residuals to the best-fit double-\texttt{CompTT} model.}
    \label{fig:spectrum}
\end{figure}

Given that the source was observed in the low-luminosity regime, we expect the spectrum to exhibit a double-hump shape.
Thus, we modeled the data with two separate Comptonization components, \texttt{CompTT} \citep{Titarchuk94} in the \texttt{XSPEC} terminology, setting a cross-normalization constant to account for calibration uncertainties between the instruments.
The resulting best-fit is shown in Fig. \ref{fig:spectrum}.
For this model, the test statistic divided by the degrees of freedom (d.o.f.) is $\chi^2$/d.o.f.$ =491/478$.
For comparison, we also tested different models in which only one continuum component is employed, modified by a discrete feature with a Gaussian absorption or emission profile (respectively, \texttt{gabs} or \texttt{Gauss} in the \texttt{XSPEC} terminology). Similar models have been tested in other XRPs, e.g., the \texttt{CompTT*gabs} model for the case of X-Persei, \citep{Doroshenko12}, the \texttt{CompTT+Gauss} model, and the \texttt{Cutoffpl*gabs} model.
The resulting $\chi^2$/d.o.f. are 495/478, 692/478, and 513/478, respectively.
However, despite being statistically comparable to the \texttt{CompTT + CompTT} best-fit model, the \texttt{Cutoffpl*gabs} model returns unphysical best-fit parameters, such as a depth of the absorption line of $346^{+148}_{-39}$ keV at $1\sigma\,$c.l., and therefore it is not considered further.
Our spectral results are reported in Table~\ref{table:spectral_xmm} for the two aforementioned models that are supported by the test statistics and the physical values of their best-fit parameters.
On the other hand, because our data do not allow us to constrain the high-energy curvature component in the double \texttt{CompTT} model, we also tested a model consisting of an absorbed blackbody with a power-law tail. This model returns a larger test statistic, $\chi^2$/d.o.f.= 521/470. Its best-fit blackbody temperature and radius are $1.32^{+0.01}_{-0.01}$\,keV and $94^{+2}_{-2}\,$m, respectively, while the photon index is $1.60\pm0.04$.
Finally, following \citet{Salganik2025}, we also tested a double \texttt{CompTT} model in which the optical depth of the high-energy \texttt{CompTT} component is frozen to $\tau_p=100$. 
The resulting test statistic is comparable to that obtained when the parameter value is let free. Also, the resulting temperature of the hot \texttt{CompTT} component is better constrained, $kT_{Hot}=9.0_{-1.2}^{+1.5}\,$keV.
It is hard, however, to reconcile such a high optical depth with a physical plasma at a temperature this high and in the low accretion rate regime. We therefore will not consider this fit as a viable model.

The value of the cross-calibration constant between \xmm\ and \nustar\ detectors differs by about $40\%$.
This is higher than the $\sim20\%$ difference expected from known calibration issues\footnote{\url{https://www.cosmos.esa.int/web/xmm-newton/sas-thread-pn-spectrum\#cav}}, implying that the \xmm\ recovered flux is systematically lower than that measured by \nustar\ by about $20\%$.
Despite the correction for the calibration issues being applied (see Sect. \ref{subsec:xmm-analysis}), we note that this correction only extends down to 3 keV for the \xmm\ response file (see also \citealt{Kang2023}).
We conclude that the difference observed in the cross-correlation constants is likely due to calibration issues and partially due to intrinsic source variability, to a factor that is at least equal to the remaining $20\%$ difference.

As a result of the different cross-calibration constant values, the source unabsorbed luminosity (in the 0.1--100\,keV range) is $5.8(1)\times10^{33}\,$erg\,s$^{-1}$ as observed by \xmm-pn, while it is $1.4(1)\times10^{34}\,$erg\,s$^{-1}$ as observed by \nustar (for an assumed distance of 3.6 kpc).
For comparison, we also report the results of the \swift/XRT observation of \src\ in 2018. The faint spectrum can be fitted either with a power-law model or a blackbody model. Given how faint the source is and the short exposure, the best-fit parameters are not firmly constrained. However, for both models, we obtain a value of the unabsorbed flux (in the 0.3–10 keV energy band) of $\sim2_{-1}^{+2}\times10^{-12}\,$erg\,cm$^{-2}$\,s$^{-1}$, corresponding to a luminosity of $\sim2\times10^{33}\,$erg\,s$^{-1}$.
We notice that this flux value is in broad agreement with that obtained by \xmm\ (see Table~\ref{table:spectral_xmm}), thus supporting the hypothesis that the source actually underwent a moderate accretion-rate change between the \xmm\ and \nustar\ observations, separated by about a week.

\begin{table}[!t]
\caption{Best-fit parameters of \src\ data from \xmm+\nustar observations for two different spectral models. All errors are reported at $1\sigma\,$c.l.} \label{table:spectral_xmm}
\centering
\begin{tabular}{lcc}
\hline\hline
\vspace{-0.2cm}
Model & \multicolumn{2}{c}{CompTT + CompTT} \\
\\
Component & Thermal Compt. & Hot Compt.\\
N$_{\rm H}$ [$10^{22}\,$cm$^{-2}$] & \multicolumn{2}{c}{$0.34^{+0.03}_{-0.03}$} \\
T$_0$ [keV] & \multicolumn{2}{c}{$0.78^{+0.06}_{-0.07}$} \\
kT [keV] & $1.8^{+0.4}_{-0.2}$ & $21.5_{-11.0}^{\star}$ \\
$\tau_p$  & $8.5^{+2.3}_{-1.9}$ & $2.8^{+3.6}_{-2.1}$ \\
Flux$^\dagger$ & $2.7^{+0.5}_{-0.3}$ & $2.7^{+0.4}_{-0.4}$\\
$\chi^2_{d.o.f.}$ & \multicolumn{2}{c}{491/478} \\
\\
\hline\\
\vspace{-0.2cm}
Model & \multicolumn{2}{c}{CompTT * gabs}  \\
\\
Component & Comptonization & Absorption line\\
N$_{\rm H}$ [$10^{22}\,$cm$^{-2}$] & \multicolumn{2}{c}{$0.41^{+0.03}_{-0.03}$} \\
T$_0$ [keV] &  $0.88^{+0.02}_{-0.02}$ & -- \\
kT [keV] & $30^{+63}_{-17}$ & --  \\
$\tau_p$  & $0.8^{+4.2}_{-0.5}$ & -- \\
E$_{Gauss}$ [keV]& -- & $18.3^{+1.2}_{-1.1}$ \\
$\sigma_{Gauss}$ [keV]& -- & $6.4^{+1.6}_{-1.2}$\\
$d_{Gauss}$ [keV] & -- & $11.8^{+6.8}_{-4.6}$\\
Flux$^\dagger$ & $3.73^{+0.05}_{-0.05}$ & \\
$\chi^2/{d.o.f.}$ & \multicolumn{2}{c}{491/466} \\
\midrule
\hline 
\end{tabular}
\tablefoot{
$^\star$Upper limit unconstrained.
\quad$^\dagger$Unabsorbed flux values calculated separately for each of the \texttt{CompTT} components in the $0.5-100\,$keV band and reported in units of $10^{-12}\,$erg\,cm$^{-2}\,$s$^{-1}$. Flux values with estimated errors were derived using the \texttt{cflux} model from \texttt{XSPEC} as obtained from the \xmm pn instrument.
}
\end{table}

\section{Timing analysis}

Given the long spin period of \src\ and the length of \nustar\ exposure, we first used \nustar\ products for the timing analysis.
We followed a procedure similar to the one presented in \citet{Pike2023} for determining the pulse period.
First, we barycentered the events using the \texttt{barycorr} tool and \nustar\ clockfile \texttt{nuCclock20100101v209}. Next, we used \texttt{Stingray} \citep{Huppenkothen2019} to perform a coarse $Z^2_n$ search \citep{Buccheri+83} with $n=4$ in the mHz range. We found one significant peak near the previously measured pulse period. Using the results of this coarse search, we next performed a fine search on the highest peak to determine the pulse period more precisely. We fitted a cubic spline to the $Z^2_4(\nu)$ distribution resulting from the fine search and determined the pulse period corresponding to the peak of this distribution, that is $P=1082.0\,$s ($\nu=0.9242\,$mHz). 

\begin{figure}[!t]
    \centering
    \includegraphics[width=0.95\columnwidth]{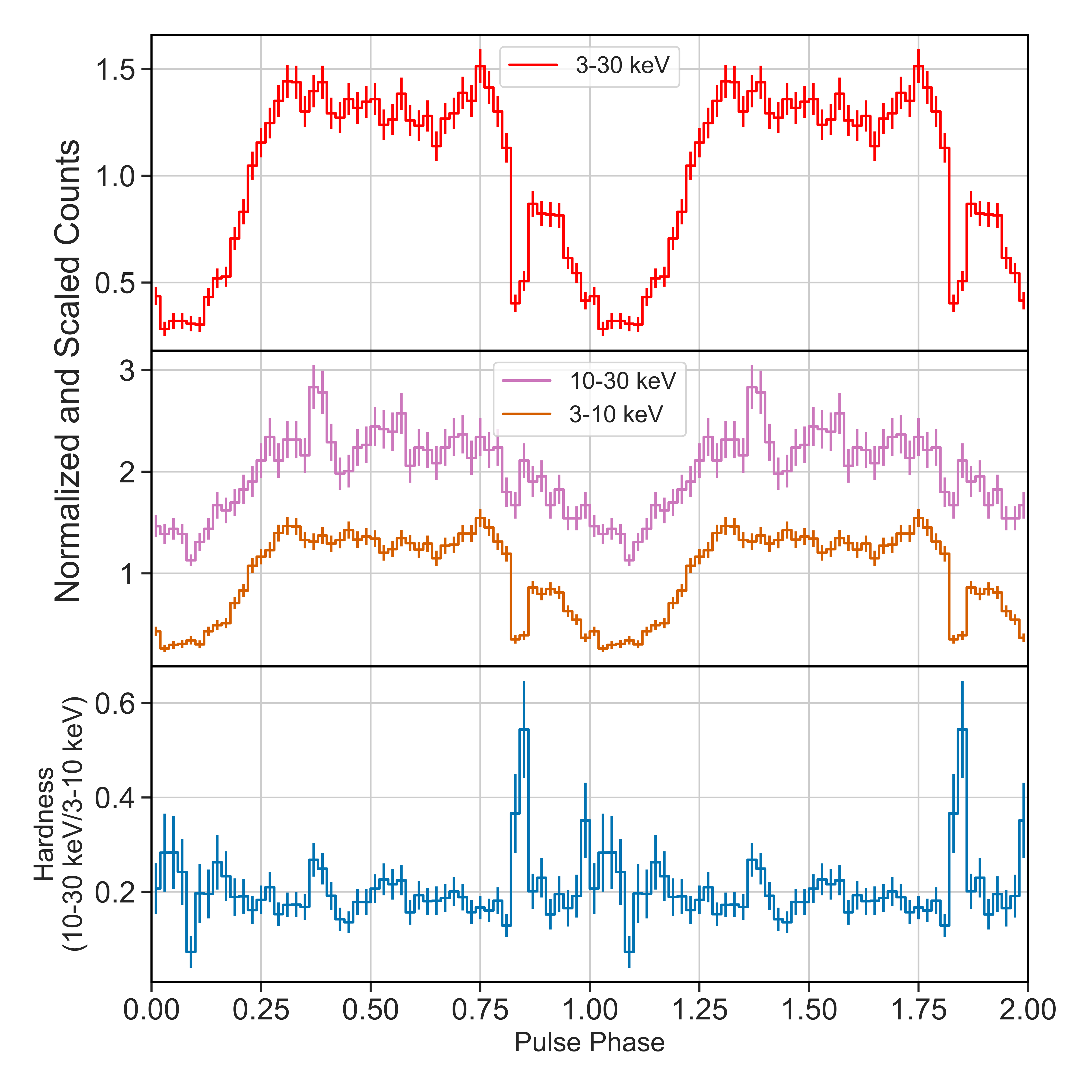}
    \caption{Top: 3-30 keV pulse profile measured by \nustar. Middle: Hard (10-30 keV, purple) and soft (3-10 keV, orange) pulse profiles, scaled and normalized for clarity. Bottom: phase-resolved hardness ratio between the hard and soft pulse profiles. 
    }
    \label{fig:pulse-profile-hardness}
\end{figure}

\begin{figure}[!t]
    \centering
    \includegraphics[width=0.95\columnwidth]{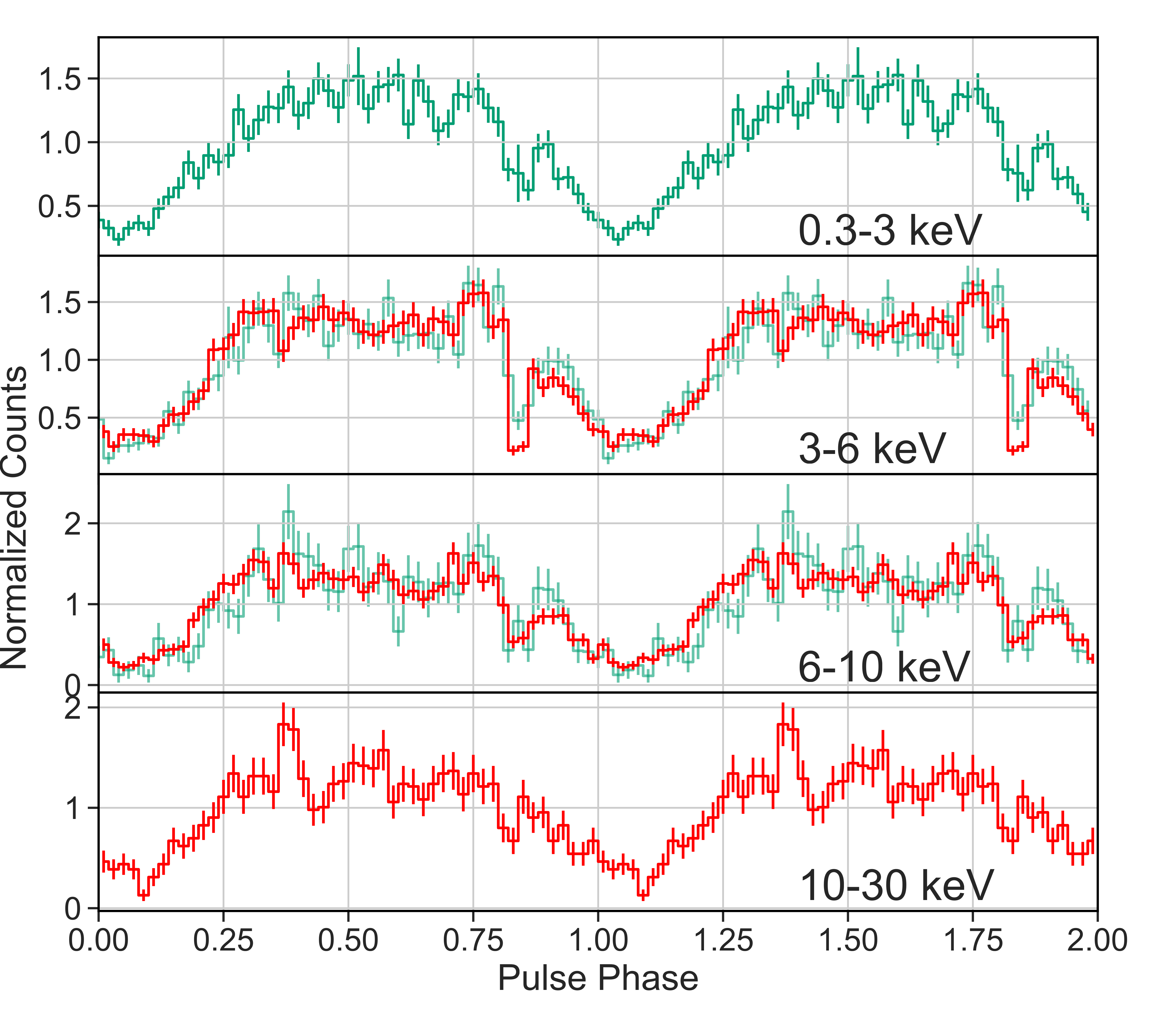}
    \caption{Energy-resolved pulse profiles measured with \nustar\ (red) and \xmm-pn (green). The pulse profile shape varies with energy. The sharp dip around $\phi=0.85$ in particular shows a strong dependence on energy, being most prominent in the 3-6\,keV range.}
    \label{fig:pulse-profiles-xmm}
\end{figure}

\begin{figure}[!t]
    \centering
    \includegraphics[width=0.95\columnwidth]{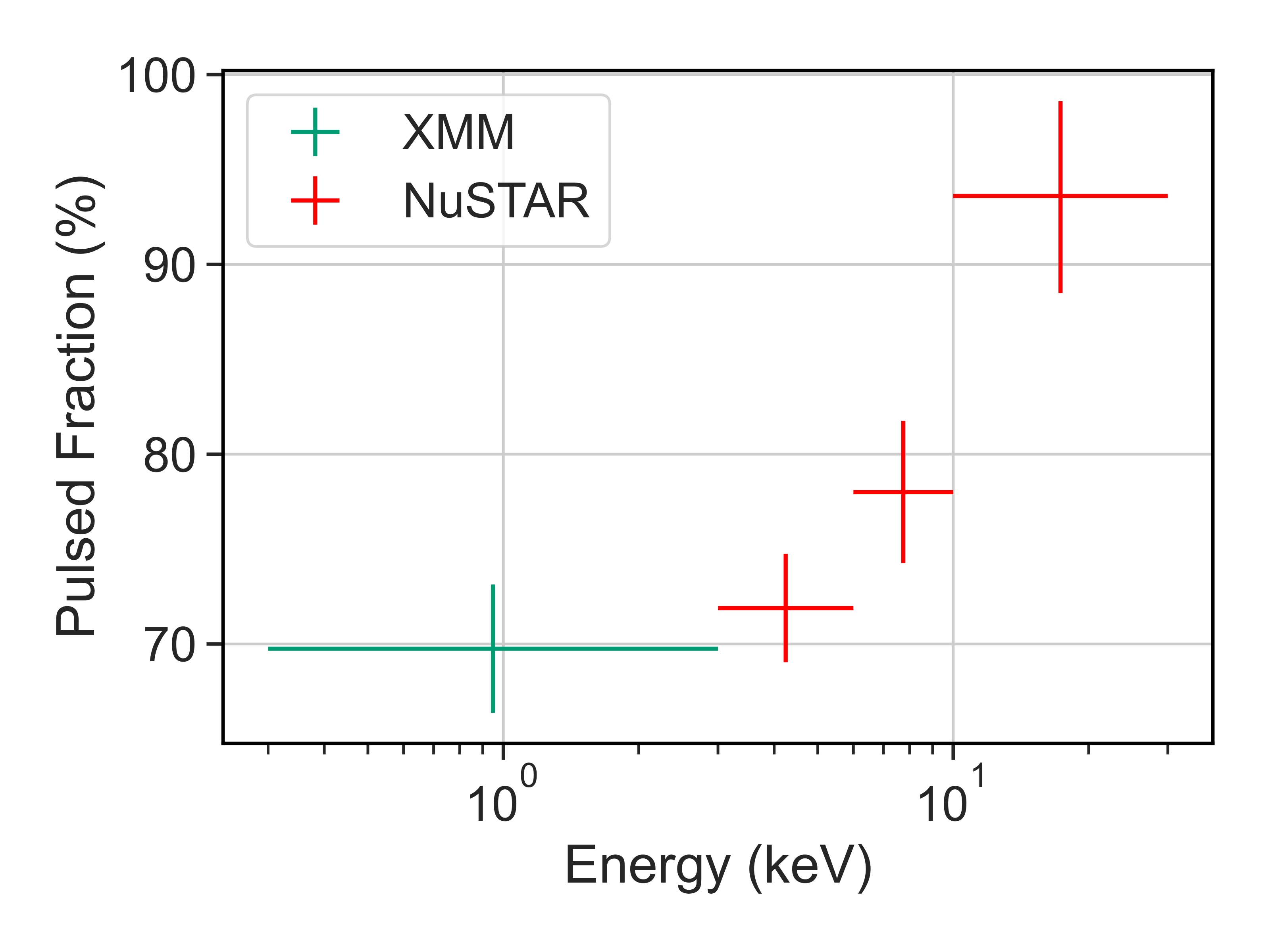}
    \caption{Energy-resolved pulsed fraction $\mathrm{PF_\mathrm{minmax}}$ as measured by \xmm (green) and \nustar (red). The pulsed fraction estimator exhibits a high overall value which strongly correlates with photon energy.}
    \label{fig:pulse-fractions}
\end{figure}

Finally, we used both \xmm\ and \nustar\ datasets together, along with a phase-fitting technique \citep{Deeter+81} to infer the best possible value of the period, which resulted in $P=1081.86\pm0.02$\,s ($\nu=0.92433(1)$\,mHz; uncertainties at 1$\sigma$ for one parameter of interest).
For this, we used a sinusoidal template profile in the overlapping energy band between the instruments and used the phase corresponding to the midway point between the maximum and the minimum values of the sinusoidal profile.
\citet{Pike2023} measured a spin period of $1085\pm1$\,s (at $90\%$ c.l.) in August 2022, following the 2022 giant outburst of \src. We therefore infer a secular spin-up during the period in between of about $-(4.6\pm1.5)\times$10$^{-8}$ s s$^{-1}$. Additionally, using Stingray, we sorted events into 50 phase bins at the measured pulse period to produce pulse profiles for four energy bins: 3-30\,keV, 3-6\,keV, 6-10\,keV, and 10-30\,keV for \nustar, and 0.3-3\,keV, 3-6\,keV, and 6-10\,keV for \xmm. 
Above 30 keV, the low flux prevents obtaining a meaningful \nustar\ pulse profile.
We show the resulting pulse profiles and the phase-resolved hardness ratio (10-30\,keV/3-10\,keV) in Figures \ref{fig:pulse-profile-hardness} and \ref{fig:pulse-profiles-xmm}. 
We also determined the energy-resolved pulsed fraction for different energy ranges, namely, 0.3-3\,keV, 3-6\,keV, 6-10\,keV, 10-30\,keV, as given by 
\begin{equation*}
\mathrm{PF_\mathrm{minmax}}=\frac{N_\mathrm{max}-N_\mathrm{min}}{N_\mathrm{max} + N_\mathrm{min}}
\end{equation*} 

where $N_\mathrm{max}$ is the value of the maximum bin for a given pulse profile, and $N_\mathrm{min}$ is the value of the minimum bin. For all pulsed fraction calculations of the $\mathrm{PF_\mathrm{minmax}}$ estimator, we rebinned the profile into 25 bins to slightly smooth the profile. In order to account for background (e.g., non-source) contributions to the pulse profile, we simulated 1000 background-only pulse profiles given the total background counts measured for each energy bin. We subtracted these simulated background profiles from the measured pulse profile, weighting the background profile according to the difference in area between the source and background extraction regions. The result is 1000 hypothetical source-only pulse profiles. For each of these profiles, we then simulated 1000 iterations assuming Poisson statistics for each bin and calculated the resulting pulsed fraction. The mean of simulated pulsed fractions for each energy bin is shown in Figure \ref{fig:pulse-fractions}, where uncertainties correspond to 1-$\sigma$ confidence regions. For the 3-30\,keV energy range, we find a pulsed fraction of $\mathrm{PF_\mathrm{minmax}}=(72^{+2}_{-3})\%$. Moreover, we find that the pulsed fraction increases with photon energy, reaching $\mathrm{PF_\mathrm{minmax}}=(94\pm5)\%$ in the hard X-ray range of 10-30\,keV.

Pulse profiles are characterized by a single-peak shape with a plateau spanning over half of the rotational phases at all energies, with the exception of a narrow dip around phase $0.85$. To investigate the dip feature, we performed phase-resolved spectroscopy, extracting \xmm\ and \nustar\ spectra within the bin $0.85-0.89$ in phase.
To fit the phase-resolved spectra, several different models can fit the data with comparable test statistic values due to the low statistics.
None of them supports a significantly larger value of N$_{\rm H}$ compared to the best-fit phase-averaged models.
However, due to the low statistics, the results were inconclusive regarding the best-fit model and its best-fit parameters, which in several instances remained unconstrained.

\section{Pulsed fraction spectra}\label{sec:pf_rms}

\begin{figure*}[t]
    \centering
    \setlength{\abovecaptionskip}{2pt} % space between figure and caption
    \setlength{\belowcaptionskip}{0pt} % space below caption
    \includegraphics[width=0.32\textwidth]{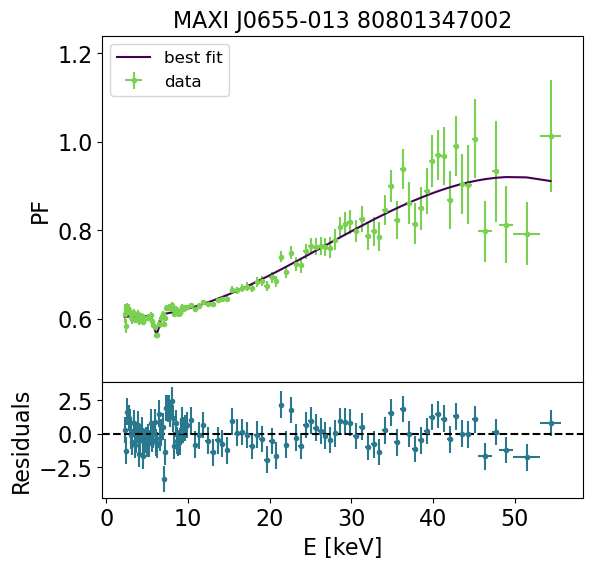}%
    \includegraphics[width=0.32\textwidth]{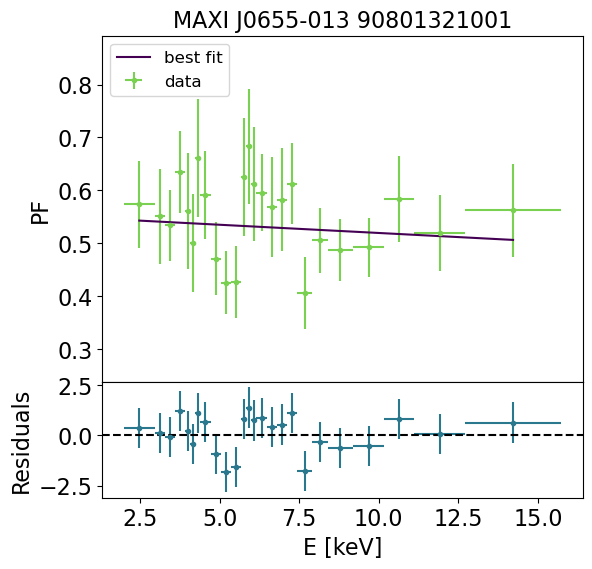}%
    \includegraphics[width=0.32\textwidth]{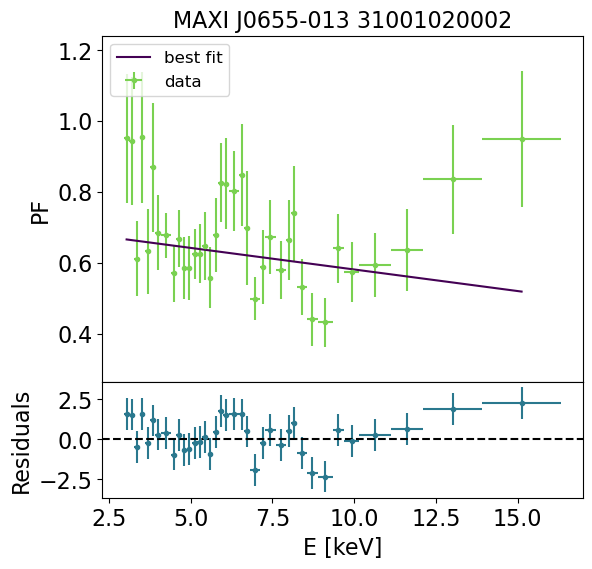}
    \caption{Pulsed fraction spectra (see Sect.~\ref{sec:pf_rms}) of the three \textit{NuSTAR} observations of \src\ taken during the 2022 outburst (ObsID 80801347002 and 90801321001), and in 2024 during the low-luminosity regime (ObsID 31001020002).
    Upper panels show data and a best-fitting polynomial fit. Lower panels
    show residuals in units of sigma with respect to the best fitting model. 
    Note for the ObsID 80801347002 the presence of an additional
    Gaussian absorption line at $\sim$\,6.2 keV.
    For the other two observations the fitted linear slope is consistent with zero at the 2$\sigma$ c.l.}
    \label{fig:pfs}
\end{figure*}

We also investigated the dependence of the pulsed fraction on energy by employing the \textit{NuSTAR}-tailored pipeline described in \citet{Ferrigno2023} \citep[see also][for recent applications]{DAi2025, Maniadakis2025}. 
This allows for optimal energy binning and computation of the pulsed fraction estimator that is not biased against extreme values or the number of bins \citep{Antia2021}.
In this pipeline, pulse profiles are initially extracted in energy intervals corresponding to the intrinsic energy resolution of the \nustar\ FPM detectors and folded using a predefined number of phase bins ($N_{\textrm{bins}}$). The profiles are then dynamically rebinned to achieve a minimum signal-to-noise ratio (S/N$\gtrsim6$) per pulse profile.

Once the rebinned profiles are obtained, the pulsed fraction is computed following the root-mean-square definition \citep[$\mathrm{PF_\mathrm{rms}}$, see Eq. A.3 in][]{Ferrigno2023}. The associated uncertainties are estimated through a bootstrap procedure, in which $10^3$ synthetic Monte Carlo profiles with the same statistical properties as the original data are generated and analyzed.
We refer to the resulting dataset of energy-resolved $\mathrm{PF_\mathrm{rms}}$ values as the pulsed fraction spectrum (PFS), owing to its similarity to energy spectra in revealing local spectral features. To search for such features, we fitted the PFS using a least-squares procedure with a polynomial function, increasing the polynomial degree until an adequate fit was obtained (see \citealt{Ferrigno2023} for more details). All polynomial coefficients were treated as free parameters.

To investigate the dependence of the $\mathrm{PF_\mathrm{rms}}$ on both energy and the accretion regime, we also applied the same analysis to two additional \textit{NuSTAR} observations previously presented in \citet{Pike2023} that sample the 2022 outburst at different stages (ObsID 80801347002 and 90801321001).
Because the energy-resolved profiles are characterized by different statistical qualities and signal-to-noise ratios (S/N), the energy range over which reliable measurements of the $\mathrm{PF_\mathrm{rms}}$ can be obtained varies among observations. To achieve an optimal balance between energy resolution and well-constrained $\mathrm{PF_\mathrm{rms}}$ values, different phase-binning schemes and S/N thresholds were adopted depending on the data quality.
For the observation with the highest statistics (ObsID 80801347002), we used 32 phase bins and a minimum S/N of 8. For the two remaining observations (ObsIDs 90801321001 and 31001020002), we employed 12 phase bins and a minimum S/N of 6. This strategy also defines the maximum usable energy range: for the lower S/N observations, PF measurements in the highest energy bins are affected by large uncertainties and were therefore excluded from the fit, limiting the PFS energy band to approximately 16 keV.

Our results are reported in Fig. \ref{fig:pfs}, where we show the PFS and the best-fitting results for the three \nustar\ observations. 
The PFS of ObsID 80801347002 shows a clear energy dependence, characterized by a smooth increase from low to high energies. No clear feature indicative of a cyclotron line is observed at 44 keV or at other energies. On the other hand, a pronounced dip is observed in the PFS around the iron line energy band. By modeling this feature with a simple Gaussian absorption profile, we obtained a best-fitting centroid energy of $6.20 \pm 0.05$ keV and a width of $0.35 \pm 0.06$ keV (1$\sigma$ c.l.). Remaining residuals following the fit suggest that the feature may have a more complex structure than that captured by a single Gaussian component. This, combined with the source falling near the detector edge, is likely the cause of the 6.4 keV iron line energy shift.

The PFS of the two low-luminosity observations exhibits similar shapes, although they can be well described only up to $\sim$20 keV. No significant energy dependence of the $\mathrm{PF_\mathrm{rms}}$ is observed within this range, as the fitted linear slope is consistent with zero at the 2$\sigma$ c.l. in both cases.

\section{Discussion}

\subsection{The double-hump spectrum}

A double-hump spectrum in XRPs has been previously observed in GX~304-1 \citep{Tsygankov2019}, A~0535+262 \citep{Tsygankov2019b}, GRO~J1008-57 \citep{Lutovinov21}, and very recently also in 2RXP J130159.6-635806 \citep{Salganik2025} during the low-luminosity regime at $\sim10^{34}\,$erg s$^{-1}$. A similar phenomenon has also been observed for X-Per at $\sim10^{35}\,$erg s$^{-1}$ \citep{Doroshenko12}. Recently, such a spectral shape has been modeled as the result of two distinct components, i.e., the low-energy ``thermal'' hump, dominated by photons with extraordinary polarization coming from deeper atmospheric layers (as they are scattered less frequently than ordinary photons), while the high-energy component is due to resonant Comptonization in the top heated layer of the atmosphere \citep{Sokolova-Lapa2021}. Based on the latter work, Zalot et al. (in prep.) are currently working on an alternative approach to estimating the magnetic field strength by using the position of the high-energy component of the double-humped spectrum.  Their preliminary estimate of the magnetic field strength is in agreement with the lower end of the range proposed by \citet{Pike2023}.

Our data do not allow us to appreciate any high-energy spectral curvature at the end of the observed bandpass (such an impediment is, in fact, encountered for most of the sources observed in the low-luminosity regime; see, e.g., GRO J1008-57, A 0535+262, 2RXP J130159.6-635806). 
On the other hand, there are cases where a double-hump spectral shape at low-luminosity is not observed altogether, that is, Swift J0243.6+6124 and MAXI J0903-531 \citep[respectively]{Doroshenko2020_J0243, Tsygankov2022_MAXIJ0903}. At least for the case of MAXI J0903-531, the absence of a double-hump spectrum at low-luminosity was interpreted in terms of a relatively low magnetic field, which would shift the high-energy bump, typically appearing around the cyclotron line centroid energy, into the bandpass of the low-energy bump, thus observationally merging with it.
We nonetheless note that, in the case of our observations of \src, the best-fit model is the one consisting of two distinct \texttt{CompTT} components, and that its best-fit spectral parameters (see Table \ref{table:spectral_xmm}) are in agreement with those observed in other sources whose data coverage or quality is such that the high-energy curvature is evident in their spectral shape (e.g., GX 304-1 and X-Per). 
The temperature of the seed photons, T$_0\approx0.8\,$keV, is in agreement with that measured for other sources in the low-luminosity regime \citep{Tsygankov2019_a0535}. The plasma temperature $kT$ of the thermal and hot components, $1.8$ keV and $21.5$ keV respectively, is also in agreement with the values obtained from the other sources of reference \citep{Tsygankov2019_a0535, Tsygankov2019}, although the value of the hot component is consistent with that found in other sources mainly due to its high uncertainty.
On the other hand, the plasma optical depth of the hot \texttt{CompTT} component tends to span an order of magnitude among the observed sources, with values that are often unconstrained and typically $\gtrsim10$. On the contrary, for \src\ we obtain $\tau_p=2.8_{-2.1}^{+3.6}$, which is consistent with the value obtained for GRO J1008-57 in the low-luminosity regime \citep{Lutovinov21}.
The plasma temperature of the Hot Comptonized component is nominally high but has an unconstrained upper limit. At face value, this parameter indicates the over-heating of the upper atmospheric layers.
In this regard, we notice that a fit with a frozen value of $\tau_p=100$ for the hot \texttt{CompTT} component (similarly to \citealt{Salganik2025}) returns a test statistic comparable to the model where the same component was allowed to vary during the fit procedure. 
%$\Delta\chi^2=2 for 1 d.o.f.$
Fixing the plasma optical depth value results in a well constrained plasma temperature of $kT_{Hot}=9.0_{-1.2}^{+1.5}\,$keV. This value is consistent with that observed in the above-mentioned works within $1\sigma$.
However, we also notice here that such a high value of the optical depth does not reflect our current understanding of the NS atmosphere, especially in the low-accretion regime.
Moreover, depending on the electron density, it is expected that a plasma with such a high optical depth will reach thermal equilibrium, where a blackbody-like spectrum forms \citep{Sokolova-Lapa2021}. 
Such a fit should therefore be treated in a purely phenomenological way.

\subsection{Discrete spectral features}

Our \xmm+\nustar\ spectral analysis shows no clear evidence of a cyclotron resonant scattering feature or an iron K$\alpha$ emission line. In contrast, those features were observed in the XRPs A~0535+262 \citep{Tsygankov2019_a0535} and GRO~J1008-57 \citep{Lutovinov21}, respectively. 
However, we find evidence of an iron line in our PFS analysis (see Sect.~\ref{sec:pf_rms} and Sect.~\ref{subsec:pp+pf} for a discussion).
\citet{Pike2023} also do not find any spectral evidence of an emission iron line, but find hints of a spectral absorption feature that they tentatively interpret as a cyclotron line at 44 keV.
In this regard, the \texttt{CompTT + gabs} model from Table~\ref{table:spectral_xmm} fits our data, but there is no evidence from previous observations of an absorption feature around 18 keV.
A centroid energy of 18.3 keV would require a harmonic ratio of 2.4 to be interpreted as the fundamental line of the absorption feature hint observed by \citet{Pike2023} at 44 keV.
Non-harmonically spaced cyclotron line energies have been theoretically expected and observed in other sources, with a factor of up to about $30\%$ difference from the $1:2$ ratio \citep{Nishimura2005, Schonherr2007, Fuerst+14}.
However, considering the luminosity-dependence shown by cyclotron line sources (see, e.g., \citealt{Staubert2007, Malacaria+15, Klochkov+11}), a luminosity change of a factor of 10 implies a change in the centroid energy of the cyclotron line of about 15\%.
On the other hand, the luminosity value measured during the observation where the hint of a cyclotron line at 44 keV was found is $\sim5\times10^{36}\,$erg s$^{-1}$ \citep{Pike2023}, while the luminosity value from our analysis is $\sim5\times10^{33}\,$erg s$^{-1}$. If the luminosity-dependence follows  the above-mentioned trend, a factor of $10^3$ in luminosity would therefore imply a shift in the cyclotron line centroid energy that is three times larger (i.e., $45\%$), which would bring the line to about 24 keV. This is significantly different from the Gaussian centroid best-fit energy of $18.3_{+2.1}^{-1.8}$ keV (at a $90\%$ c.l.). 
Moreover, we notice that the width and depth that we measure for the absorption feature ($6.4_{-1.2}^{-1.6}$ keV and $11.8_{-4.6}^{+6.8}$ keV, respectively) are typical of cyclotron lines \citep{Staubert19}, although the large errors are likely indicative that the model is not appropriate.
For these and more reasons (see Sect.~\ref{subsec:timing_res} and \ref{subsec:pp+pf}), we disfavor the interpretation of the \texttt{CompTT*gabs} spectral model as an indication of the presence of a cyclotron line in \src.

\subsection{Timing results}\label{subsec:timing_res}

In our low-luminosity observations of \src\ we measured a spin period of about 1081.86$\pm0.02$ s. This is lower than the spin period measured by \citet{Pike2023} of $1085\pm1$ s (at $90\%$ c.l.) obtained during the giant outburst in 2022. 
The spin derivative between those two measurements is $-(4.6\pm1.5)\times$10$^{-8}$\,s\,s$^{-1}$. %($\dot{\nu}=3.7\pm1.3\times10^{-14}\,$Hz\,s$^{-1}$).
This implies that the source has been spinning-up since the last outburst.
This is consistent with the interpretation of the source belonging to the class of persistent BeXRBs \citep{Reig1999}, characterized by low persistent X-ray luminosity, long ($\gtrsim200\,$s) spin periods, and absent or weak iron emission lines in their X-ray spectrum \citep{Reig2011}.
In this regard, we point out that \src\ had been previously detected by SRG/ART-XC \citep{Sunyaev2021, Pavlinsky2022, Sazonov2024} at a 4–12 keV flux of $\sim7\times10^{-12}$ erg cm$^{-2}$ s$^{-1}$ ($L_{\mathrm{X}}\sim10^{34}$ erg s$^{-1}$), indicating that the source was already active at low luminosity outside the outburst.
Moreover, assuming balance between the accretion torque $\dot{M} \sqrt{G\,M_{NS}\,R_{cor}}$ at the co-rotation radius, $R_{cor}$, and the angular momentum evolution, $I\dot{\Omega}=I\,2\pi\dot{\nu}$, then the measured spin period derivative corresponds to a value of the X-ray flux due to residual accretion of about $\sim3\times10^{-11}$ erg cm$^{-2}\,$s$^{-1}$. At a distance of 3.6 kpc, this corresponds to a luminosity of $\sim4\times10^{34}$ erg\,s$^{-1}$, which is larger than the luminosity observed in this work by \xmm or \nustar by a factor of a few. This tension may indicate that the source has experienced strong sporadic spin-up episodes that went unobserved in the period between the outburst in 2022 and our observations in 2024.

During the 2022 outburst, \citet{Pike2023} measured a local spin-up trend that they interpreted according to the torque accretion model by \citet{GLb}. 
This allowed them to recover a magnetic field strength of about $5\times10^{14}\,$G. 
By applying the same method, we obtain, for the same magnetic field strength and employing the luminosity value measured by \xmm-pn, a spin-up value of about $-6.4\times$10$^{-6}$\,s\,s$^{-1}$. This is about two orders of magnitude stronger compared to the value inferred above as the spin derivative between the end of the outburst in 2022 and our observations in 2024.
Magnetic field measurements through torque models are, however, troubled by systematic uncertainty of the physical details of how matter couples to the magnetic field. The difference between the models of how the predicted spin-up is related to the magnetic field value becomes even more significant for low mass accretion rates \citep{Stierhof2025}. In this regime, simulations also show a deviation from the torque models, indicating a change in the coupling mechanism not captured by the models \citep[e.g.,][but scaled for T\,Tauri stars]{Ireland2022}.
This could explain the high value obtained for the magnetic field, which would further imply a large magnetospheric radius, with consequences for the accretion dynamics.
In fact, the rotating magnetosphere acts as a centrifugal barrier once the mass accretion rate drops below a certain value, thus settling the source in the so-called propeller regime \citep{Illarionov75}.
The luminosity corresponding to the onset of the centrifugal barrier is \citep{Stella1986, Campana2002}:
\begin{equation}\label{eq:propeller}
L_{\text{prop}}\approx4\times10^{37}\,k^{7/2}\,B_{12}^2\,P^{-7/3}\,M_{1.4}^{-2/3}\,R_6^5\, \text{erg}\,\text{s}^{-1}
\end{equation}
where $M_{1.4}$ is the neutron star mass in units of $1.4M_\odot$, R$_6$ is the NS radius in units of $10^6$ cm, $P$ is the NS spin period in seconds, B$_{12}$ is the magnetic field strength in units of $10^{12}$ G, while $k$ is the coupling factor relating the magnetospheric radius for disk accretion to the Alfvén radius calculated for spherical accretion (usually assumed to be $k=0.5$). Employing the above-mentioned value of the magnetic field strength and a spin period of 1082 s, a canonical NS mass of 1.4 M$_\odot$ and a NS radius of 12 km, we obtain a propeller luminosity of about $\sim1.6\times10^{35}\,$erg s$^{-1}$.
On the other hand, if the magnetic field strength value is derived from the supposed cyclotron line at 44 keV, then we obtain a propeller luminosity of about $\sim1.8\times10^{31}\,$erg s$^{-1}$.

These measurements span a wide range of luminosity values, also covering those obtained by \swift/XRT, at $\sim2\times10^{33}\,$erg s$^{-1}$. The latter was obtained from a best-fit spectrum that is consistent with a power-law, although a blackbody fit cannot be ruled out due to the poor statistics. If the luminosity observed by \swift/XRT was actually due to a power-law spectrum, it would be considered an indication of residual accretion \textbf{(see, e.g., \citealt{Tsygankov2016})}, thus implying that the source has not yet entered the propeller regime. However, given that the available statistics prevent us from drawing such a conclusion, and because we cannot estimate the pulsating nature of the source at the time of the \swift/XRT observation, we conservatively adopt the \xmm-pn luminosity ($L_{\text{XMM}}\sim5.8\times10^{33}\,$erg s$^{-1}$, see Sect.~\ref{sec:spectral_analysis}) as the upper limit for entering the propeller regime (that is $L_{\text{prop}}\lesssim\,L_{\text{XMM}}$). This, in turn, allows us to put an upper limit on the magnetic field strength (see Eq.~\ref{eq:propeller}) as $B\lesssim8.9\times10^{13}\,$G.

\subsection{Pulse profiles and pulsed fraction}\label{subsec:pp+pf}

We obtained energy-resolved pulse profiles for \src\ as observed at the low-luminosity stage. These differ from the pulse profiles observed during the outburst \citep{Pike2023}. In fact, during the 2022 outburst, the pulse profiles observed at an unabsorbed source luminosity of $\sim6\times10^{36}\,$erg s$^{-1}$ showed a double- or even triple-peaked shape at energies below 10 keV and a single-peaked structure at higher energies. On the other hand, at a lower luminosity of $\sim7\times10^{34}\,$erg s$^{-1}$ the pulse profiles became simpler at all energies yet showed signs of a double-peaked structure.
The observed pulsed fractions at the two different luminosity stages were measured (for the full 3–78 keV energy range) as $84.4\%\pm0.7\%$ and $63\%\pm6\%$, respectively. 
On the contrary, here we observe a pulse profile that is overall single-peaked at all energies, with a plateau structure spanning roughly $50\%$ of the rotational phase.
Additionally, we observe a short and sharp drop in counts around phase $\phi=0.85$ which is mostly apparent in soft X-rays. The dip is coincident in phase with a spike in the hardness ratio between 3-10\,keV and 10-30\,keV energy range.

The sharp phase- and energy-dependent dip around $\phi=0.85$ is similar to a feature in the pulse profile of EXO~2030+375 reported by \citet{Ferrigno2016, Fuerst2017} and later observed again by \citet{Thalhammer2024}. \citet{Ferrigno2016, Fuerst2017} attributed this feature to increased absorption due to the self-occultation of the emitting region by part of the accretion column. 
\citet{Caballero2013} also observe a dip in the energy-resolved pulse profile of A~0535+26 below $\sim20$ keV during the decay of an outburst, at a 20-100 keV luminosity of about $3\times10^{35}\,$erg s$^{-1}$.
A similar feature was also observed in the soft energy pulse profile of the XRP RX J0440+4431 \citep{Usui12} and was likewise interpreted as being due to absorption from the accretion stream intercepting the line of sight during the NS rotation \citep{Galloway2001, Malacaria2024, Salganik2023}.
Following this scenario, we may estimate the size of the occluding structure. Given a canonical neutron star radius of 12\,km and a duration of the dip of about 4\% of the spin period, we infer that the occluding structure size, if it consists of a region projected onto the NS, would be about $3.0$\,km.
However, we also notice that, given the observed low-luminosity stage, it is unlikely that a column structure dense enough to produce the observed dip is present. 
Moreover, if the dip were solely due to absorption from neutral matter, then it would appear more pronounced at lower energies compared to the harder X-ray bands explored.
This is not the case, as shown by the comparison of the pulse profiles in Fig. \ref{fig:pulse-profiles-xmm}, where the dip appears most prominently in the 3-6\,keV band instead of the 0.3-3\,keV band.
A possible influencing factor can be the presence of partially ionized material, which is characterized by only a weak dependence on photon energy and can be approximated by the Thomson scattering cross section.
Alternatively, the soft ($\lesssim3\,$keV) emission is originating from an unabsorbed region.

Although no timing results are reported for most of the above-mentioned sources observed in the low-luminosity regime, for GRO J1008-57 the 3-79 keV pulsed fraction (measured as $\mathrm{PF_\mathrm{minmax}}$) observed at $\sim4\times10^{34}\,$erg\,s$^{-1}$ is rather low, i.e., on the order of $20\%$ \citep{Lutovinov21}.
For the persistent low-luminosity source X-Per, we also infer a $\mathrm{PF_\mathrm{minmax}}$ of about $50\%$ in the 4-11 keV range from the \textit{RXTE} pulse profiles obtained by \citet{Coburn2001}.
\cite{Jaisawal2021} also observed pulsations from EXO 2030+375 at a luminosity of about $2.5\times10^{35}$ erg\,s$^{-1}$, but with a relatively low $\mathrm{PF_\mathrm{minmax}}$ of less than $20\%$ up to 80\,keV.
However, for \src\ we obtain an energy-dependent $\mathrm{PF_\mathrm{minmax}}$ that goes up to about $100\%$ in the 10-30 keV energy band.
Such a high pulsed fraction is observed in some sources only in the harder energy bands ($\gtrsim50$ keV, \citealt{Lutovinov09}). Among them, we mention a few exemplary cases, such as SMC~X-3 \citep{Tsygankov17}, RX~J0440.9+4431 \citep{Liu2023}, 2S~1417-624 \citep{Liu2024}. 
However, although the luminosity-dependence of the pulsed fraction in those sources is not clear, we notice that all of them show high values of the pulsed fraction at luminosities that are relatively high (greater than $\sim10^{36}$ erg s$^{-1}$) or even super-Eddington, and therefore at a substantially different accretion regime.
On the contrary, there is no clear indication of any XRPs showing a $\sim100\%$ $\mathrm{PF_\mathrm{minmax}}$ in the energy range and luminosity regime considered in this work. It is not clear how such a high pulsed fraction can be produced. A possible qualitative scenario includes the idea that the hot spots responsible for the observed radiation are geometrically positioned in such a way that, at certain rotational phases, their emission is fully eclipsed from the observer line of sight. This is facilitated in the case, e.g., of a single spot emission.
However, we also notice that the flat top pulse profile is not typical of single spot emission, thus implying that more complicating factors are at work.

The PFS obtained via the $\mathrm{PF_\mathrm{rms}}$ evolution with energy and luminosity highlights several spectral properties. First, a Gaussian feature corresponding to the iron K$\alpha$ line at 6.4 keV has emerged at high luminosity. This is interesting, as persistent BeXRBs are characterized by weak or absent iron emission lines in their energy spectra. However, PFS has demonstrated the capability to uncover features that are hidden in the energy spectra (see, e.g., Ambrosi et al. subm.).
The same line is not present in the PFS retrieved at lower luminosity, although low statistics prevent us from drawing firm conclusions.
Moreover, PFS is a powerful tool for characterizing cyclotron lines by constraining both the centroid energy and the width of such features. Nonetheless, our analysis of the PFS does not show clear hints of the possible cyclotron line feature indicated by \citet{Pike2023} at 44\,keV. This supports the interpretation that such a feature is, in fact, absent in the energy spectrum.
The $\mathrm{PF_\mathrm{rms}}$ energy-dependence shows a flat trend (around $60\%$), which is contrary to what is observed from \src\ and other sources at higher luminosity. This might be due to the fact that photons from all probed energies originate in the same region and undergo the same pulsating mechanisms.
Finally, we also notice that no appreciable features are present in the PFS obtained from our low-luminosity regime observation around the 3-6 keV energy range, where the pulse profile dip is more pronounced.

\section{Conclusions}

We observed the accreting XRP \src\ during its low-luminosity accretion regime with \xmm\ and \nustar.
We obtained broadband spectra and energy-resolved pulse profiles for the first time in this regime for this source.
Our main findings can be summarized as follows:

   \begin{enumerate}
      \item The joint \xmm+\nustar spectrum is best fitted with a double-hump model. This is consistent with predictions that, at low luminosity, foresee the first hump peaking around 5 keV due to extra-ordinary polarization escaping from the lower layers of the atmosphere, while the higher energy hump peaks around 30 keV due to Comptonization by the top layers of the atmosphere over-heated by the residual accretion.
      \item We measure a spin period of $1081.86\pm0.02\,$s. This implies a spin-up compared to observations of \src\ taken in 2022 during an outburst episode. Therefore, the source has been accreting since then, instead of entering the propeller regime, thus supporting the identification of the source as a persistent BeXRB, and yielding an upper limit for the magnetic field strength of $B\lesssim9\times10^{13}\,$G.
      \item We obtain energy-resolved pulse profiles and observe an overall single-peaked profile with a plateau spanning roughly half of the rotational phase. This is markedly different from the profiles observed during the high-luminosity state obtained during the 2022 outburst. Moreover, the profiles show a narrow, sharp dip coincident with a spike in the hardness ratio, thus ascribable, at least in part, to increased absorption.
      \item We measure different estimators of the pulsed fraction. The $\mathrm{PF_\mathrm{minmax}}$ shows an energy-dependent pulsed fraction that rises to about $100\%$ in the 10-30\,keV band. This is not observed in other sources, especially in such a low-luminosity regime. We also obtain $\mathrm{PF_\mathrm{rms}}$ and pulsed fraction spectra for \src\ at different luminosity stages.
      These show a flat energy-dependence at low luminosity, and the PFS obtained at high luminosity indicates the presence of an iron K$\alpha$ emission line at 6.4 keV.
      \item Detection of a cyclotron line is not supported by our spectral analysis, nor by the \nustar\ PFS obtained at high luminosity.
   \end{enumerate}

\begin{acknowledgements}
CM acknowledges funding from the Italian Ministry of University and Research (MUR), PRIN 2020 (prot. 2020BRP57Z) ``Gravitational and Electromagnetic-wave Sources in the Universe with current and next generation detectors (GEMS)'' and the INAF Research Grant ``Uncovering the optical beat of the fastest magnetised neutron stars (FANS)''.
LD acknowledges funding from the Deutsche Forschungsgemeinschaft (DFG, German Research Foundation) - Projektnummer 549824807. RA acknowledges financial support from INAF through the grant ``INAF-Astronomy Fellowships in Italy 2022 - (GOG)''
\end{acknowledgements}

% WARNING
%-------------------------------------------------------------------
% Please note that we have included the references to the file aa.dem in
% order to compile it, but we ask you to:
%
% - use BibTeX with the regular commands:
%   \bibliographystyle{aa} % style aa.bst
%   \bibliography{Yourfile} % your references Yourfile.bib
%
% - join the .bib files when you upload your source files
%-------------------------------------------------------------------

\bibliographystyle{yahapj}
\bibliography{references}

\end{document}